\documentclass[useAMS,usenatbib]{mn2e}

\usepackage[dvips]{graphicx}
\usepackage{amssymb}
\usepackage{txfonts}

\input epsf.sty
\title[Masses of the components of the HDE 226868/Cyg X-1 binary system]{Determination of the masses
of the components of the HDE 226868/Cyg X-1 binary system}
\author[J. Zi\'o{\l}kowski]{J. Zi\'o{\l}kowski \thanks{E-mail:
jz@camk.edu.pl}\\N. Copernicus Astronomical Center, ul. Bartycka
18, 00-716 Warsaw, Poland}
\begin{document}

\date{Accepted 0000 December 00. Received 0000 December 00; in original form 0000 December 00}

\pagerange{\pageref{firstpage}--\pageref{lastpage}} \pubyear{2013}

\maketitle

\label{firstpage}

\begin{abstract}
Recent determination of the distance to HDE 226868/Cyg X-1 binary
system (Reid et al., 2011) and more precise determination of the
effective temperature of HDE 226868 (Caballero-Nieves et al., 2009)
permit a more accurate estimate of the masses of both components.
Using up to date evolutionary models, I obtain a mass range of
between 25 to 35 M$_\odot$ for the mass of the supergiant and
between 13 to 23 M$_\odot$ for the mass of the black hole. Accepting
more liberal estimates of uncertainties in both the distance and the
effective temperature, one may extend these ranges to 21 to 35
M$_\odot$ and 10 to 23 M$_\odot$ for both masses, respectively. The
most likely values within these ranges are, respectively, 27
M$_\odot$ and 16 M$_\odot$. The obtained mass of black hole agrees
with the value 15 $\pm$ 1 M$_\odot$ suggested by Orosz et al.
(2011). However, the value suggested by them for the mass of the
supergiant of 19 $\pm$ 2 M$_\odot$ should not be used as such a star
violates the mass-luminosity relation for the the massive core
hydrogen burning stars. This consideration was not incorporated into
the iterative process of Orosz et al.

To resolve this violation I consider the possibility that the
hydrogen content of HDE 222268 might be lowered as a result of the
mass transfer and the induced fast rotation of the mass gainer. I
analyzed the evolutionary effects of such situation and found that,
while important, they do not invalidate the conclusions listed
above. If, as a result of the rotation induced mixing, the present
hydrogen content of HDE 226868 is equal about 0.6 (as suggested by
some observational data), then its present mass may be somewhat
lower: $\sim$ 24 M$_\odot$ rather than $\sim$ 27 M$_\odot$.

\end{abstract}


\section{Introduction}

Cyg X-1 was the first recognized black hole (Bolton 1972) but the
masses of the binary components are still a subject of controversy.
The long history of their determinations prior to 2005 is given by
Ziolkowski (2005). After 2005, two major observational improvements,
crucial for mass determination, took place. First, Caballero-Nieves
et al. (2009, hereafter C-N+09) through careful stellar atmosphere
modeling obtained more precise estimate of the effective temperature
of supergiant HDE 226868, which is a binary companion of Cyg X-1.
Second, Reid et al. (2011, hereafter R+11) estimated (through a
radio parallax) the distance to the binary system HDE 226868/Cyg
X-1. Orosz et al. (2011, hereafter O+11) used this distance to
estimate the parameters of the binary system. They have done a very
precise job, using 8 free parameters and fitting almost 600
observables: 3 UBV light curves (20 points each) and radial velocity
curve (529 points). The values of the parameters were determined
through an iterative scheme. The additional free parameter was the
effective temperature of the supergiant. This parameter, however,
was not iterated with the others, but was adjusted separately (the
preferred value was found to be $T_{\rm e}$ = 31 000 K). Allowing
for non-synchronous rotation of the supergiant and slightly
eccentric orbit, O+11 were able to obtain impressively good fits of
all three light curves and of the radial velocity curve. From their
iterative process, they got $i = 27.06 \pm 0.76^{\rm o}$ for the
binary orbit inclination and $M_{\rm opt} = 19.16 \pm 1.90$
M$_\odot$ and $M_{\rm X} = 14.81 \pm 0.98$ M$_\odot$ for the masses
of both components. The formal errors are impressively low. However,
the solution has a flaw: the mass of the optical component is
inconsistent with its calculated luminosity (log $L/$L$_\odot$ =
5.352). The mass and the luminosity of massive core hydrogen burning
stars are not independent parameters, as I will demonstrate in this
paper. I will also discuss the consequences of this fact.

\section{The  Evolutionary Calculations for HDE 226868}

I calculated new evolutionary models for the supergiant HDE 226868,
making use of a new more precise estimate of its effective
temperature (C-N+09) and of the just determined distance to the
binary system (R+11). The evolutionary tracks were computed for core
hydrogen-burning phase of stars with initial masses in the range
$25$--$40$ M$_\odot$. The Warsaw evolutionary code described by
Ziolkowski (2005) was used. The standard Population I chemical
composition of $X$=0.7 and $Z$=0.02 was adopted as the initial
chemical composition.

The calculations were carried out under the assumption that the
evolution starts from homogeneous configurations (i.e. the
evolutionary clock was reset at the end of the mass transfer and the
evolution started anew). Ziolkowski (2005) gave the arguments
indicating that this is a good approximation. To calculate the
stellar wind mass loss, I applied the formula derived by Hurley,
Pols \& Tout (2000, hereafter HPT), based on parametrization of
Nieuwenhuijzen \& de Jager (1990). As in earlier papers, I
introduced the multiplicative factor $f_{\rm SW}$ applied to the HPT
formula. The evolutionary calculations indicate, that to account for
the evolutionary state of HDE 226868, with the present day
observational data, one has to use the value of $f_{\rm SW}$ in the
range 2 to 5 (the values in Tab. 1 cover the range of 2 to 8, but I
do not consider the the case with $f_{\rm SW}$ = 8 as a realistic
model $-$ see the further discussion).

\subsection{The mass-luminosity relation for massive core hydrogen
burning stars}

This relation, based on our evolutionary models, for the value of
the effective temperature $T_{\rm e}$ = 31 000 K (i.e. the value
adopted by O+11) is shown in Fig. 1. As one can see, this relation
is quite tight and the dependence on the uncertain parameter $f_{\rm
SW}$ is weak. Please, note that the value of the supergiant mass
suggested by O+11 (19.2 M$_\odot$) for their value of its luminosity
(log $L/$L$_\odot$ = 5.352) lies out of the frame of the picture.

\begin{figure}
\hbox{\epsfxsize=1\hsize\epsfbox{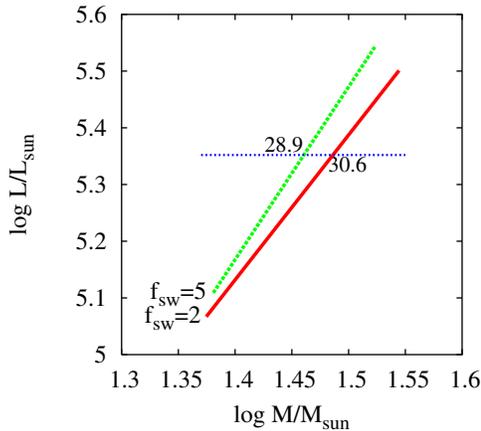}}
      \caption[h]{The mass-luminosity relation (based on the evolutionary models)
for massive core hydrogen burning stars at effective temperature
$T_{\rm e}$ = 31 000 K. The relation is given for two values of the
multiplying factor $f_{\rm SW}$ applied to HPT formula. The dotted
horizontal line corresponds to the luminosity favored by O+11
solution. The values of the masses corresponding to the crossings of
this line with $M-L$ relations are also shown (in solar units). Note
that the value of the optical component mass suggested by O+11 (19.2
M$_\odot$) lies out of the frame of the picture.}
     \label{f1}
    \end{figure}

\subsection{The evolutionary tracks in the Hertzsprung-Russell (HR)
diagram}

The observational constraints given by C-N+09 for the effective
temperature of HDE 226868 and by R+11 for the distance to the binary
system produce an error box in the HR diagram, that is shown as a
parallelogram in Fig. 2. The "best" observational solution lies in
the center of this parallelogram. The solution advocated by O+11 is
shown as a green cross. The procedure of obtaining the different
evolutionary models of HDE 226868 was the following one. For each
pair of the chosen values of $T_{\rm e}$ and $d$ (corresponding e.g.
to the center or to a corner of the error box) a corresponding
position (i.e. a pair of the values of $T_{\rm e}$ and $L$) was
calculated. Then an evolutionary track passing through that position
was calculated. The value of $f_{\rm SW}$ was adjusted so as to
obtain a stellar wind strength at this position consistent within 10
\% with the observed value ($\dot{M} = - 2.6 \times 10^ {-6}$
M$_\odot$/yr, Gies et al. 2003).

The observational estimate of the stellar wind mass flux requires a
brief comment. Stellar wind from HDE 226868 is rather complicated
(detailed modeling was done by Gies et al. 2008). It is composed of
a narrow stream of focused wind along the axis joining both
components and a roughly isotropic wind from the hemisphere which is
in the X-ray shadow (the side which is not illuminated by the X-ray
from the vicinity of black hole). The first component determines, in
first approximation, the amount of the accreted matter and so also
the X-ray luminosity. The second one determines the total mass loss
from the supergiant. From the point of view of comparing
evolutionary models with the observations, the second component is
relevant. Paper of Gies et al. (2008) does not give any quantitative
estimate. Earlier paper (Gies et al. 2003) gave $\dot{M} = - 2.6
\times 10^ {-6}$ M$_\odot$/yr for the low/hard state and $\dot{M} =
- 2.0 \times 10^ {-6}$ M$_\odot$/yr for the high/soft state. Cyg X-1
spends most of the time in the low/hard state although this time
fraction decreased recently from $\sim$ 90 \% to $\sim$ 66 \% (Wilms
et al. 2006). Moreover, Zdziarski et al. (2011) found that the
average X-ray luminosity did not change meaningfully during the hard
states of recent $\sim$ 15 years, which might imply that also the
average total mass loss was constant. Taking all this into account,
I decided to take $\dot{M} = - 2.6 \times 10^ {-6}$ M$_\odot$/yr as
the evolutionary parameter describing the present state of HDE
226868.

The procedure described at the beginning of this (2.2) section
produced several possible evolutionary models of HDE 226868 that are
listed in Tab. 1. The corresponding evolutionary tracks in the H-R
diagram are shown in Fig. 2. As one may see, all tracks predict,
with a wide margin, that HDE 226868 is in core hydrogen burning
phase. Indeed, the central hydrogen content for models in Tab. 1 is
in the range 0.13 to 0.17. This indicates that the star is in slow,
nuclear phase of its evolution and is consistent with the observed
constancy of the orbital period of the system (Gies \& Bolton 1982).

As may be seen from Tab. 1 and Fig. 2, assuming wide margins for
uncertainty of both $T_{\rm e}$ and $d$, one gets the mass of HDE
226868 in the range 20.7 to 35.4 M$_\odot$. However, I believe that
the value of $T_{\rm e}$ equal 25 500 K is rather too low for HDE
226868. Also, the value $f_{\rm SW}$ = 8, required to get a fit for
this temperature, seems to be excessively high. Therefore, I would
rather discard the third line in Tab. 1 and state that, on
evolutionary ground, the mass of HDE 226868 is probably in the range
25 to 35 M$_\odot$.

\begin{table}

\centering
 \begin{minipage}{75mm}
  \caption{Present mass of the optical component for the different assumed
values of its effective temperature and of the distance to the
   binary system. The most likely model is typed in boldface (it corresponds to the effective temperature estimate of C-N+09 and distance estimate of
R+11.}

\vbox{
\begin{tabular}{|r|r|r|r|l|l|}
\hline
&&&&&\\
\multicolumn{1}{|c|}{$d$}&\multicolumn{1}{|c|}{$T_{\rm
e}$}&\multicolumn{1}{|c|}{log$L$}&\multicolumn{1}{|c|}{$M_{\rm
opt}$}&
\multicolumn{1}{|c|}{$f_{\rm SW}$}&\multicolumn{1}{|c|}{$M_0$}\\
\multicolumn{1}{|c|}{[kpc]}&\multicolumn{1}{|c|}{[10$^3$
K]}&\multicolumn{1}{|c|}{[L$_{\odot}$]}&\multicolumn{1}{|c|}{[M$_{\odot}$]}&&\multicolumn{1}{|c|}{[M$_{\odot}$]}\\
&&&&&\\
\hline

{\bf 1.86}&{\bf 28.0}&{\bf 5.309}&{\bf 27.2}&{\bf 3.6}&{\bf 33}\\

1.98&30.5&5.513&35.4&1.76&40\\

1.75&25.5&5.094&20.7&8&29\\

1.75&28.0&5.257&25.5&4.9&32.5\\

1.86&31.0&5.352&29.4&4&36\\

\hline

\end{tabular}}

\vspace{4mm}
{\footnotesize NOTES:\vspace{2mm}\\
$f_{\rm SW}$ denotes the multiplying factor applied to HPT formula;
$M_0$ denotes the initial (ZAMS) mass of the optical component for a
given evolutionary track; other symbols have their usual meanings\\
} \vspace{0mm}
\end{minipage}
\end{table}

\begin{figure}
\hbox{\epsfxsize=1\hsize\epsfbox{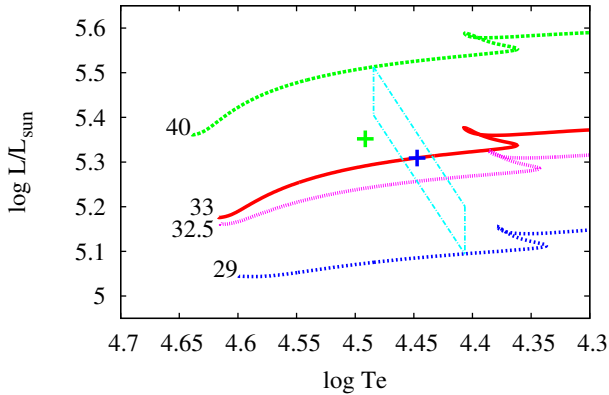}}
      \caption[h]{The evolutionary tracks in the H-R diagram. The tracks are
labeled with the initial mass of the star (in solar units). The
values of the multiplying factor $f_{\rm SW}$ applied to HPT formula
were adjusted individually for each track (see Table 1), so as to
obtain agreement with the observed value of the present strength of
stellar wind from HDE 226868. The tracks were meant to reproduce the
observed values of the present luminosity and effective temperature
of HDE 226868 according to the different solutions listed in Table
1. The parallelogram describe the observational error box using
effective temperature estimates of C-N+09 and distance estimates of
R+11 (the central values of these estimates are marked by the blue
cross). The green cross corresponds to O+11 solution.}
     \label{f2}
    \end{figure}

\section{The mass of the black hole component}

Once the evolutionary model of the supergiant component is selected
from Tab. 1, we can use two equations discussed by Ziolkowski
(2005). One of these equations makes use of the mass function,
\begin{equation}
f(M_{\rm X}) = M_{\rm opt} {\rm sin}^3 i/[q(1+q)^2].
\end{equation}
where $q = M_{\rm opt}/M_{\rm X}$ is the ratio of the
masses of the components and $i$ is the inclination of the orbit. The other relates the radius of the star to the size of
the orbit,
\begin{eqnarray}
R_{\rm opt}& =& R_{\rm RL} \times f_{\rm RL} =\nonumber\\ &=& f_{\rm
RL} (0.38 + 0.2 \log\hspace*{.5ex} q) A =\nonumber \\ &=& f_{\rm RL}
(0.38 + 0.2 \log\hspace*{.5ex} q) a_1 (1+q),
\end{eqnarray}
where $R_{\rm RL}$ is the radius of the Roche lobe around
HDE 226868, $f_{\rm RL}$ is the fill-out factor ($f_{\rm RL} =
R_{\rm opt}/R_{\rm RL}$), $A$ is the orbital separation of the
binary components and $a_1$ is the radius of the orbit of HDE 226868
around the mass center of the system.

Measurements of the radial velocities of the optical component give
us $f(M_x) = 0.251$ M$_\odot$ and $a_1$sin$i = 8.36$ R$_\odot$ (Gies
et al. in 2003). Inserting these observational data, eqs. (1)$-$(2)
can be written as:
\begin{eqnarray}
\lefteqn{
M_{\rm opt} {\rm sin}^3 i/[q(1+q)^2] = 0.251,}\\
\lefteqn{
R_{\rm opt} = f_{\rm RL} (0.38 + 0.2 \log \hspace*{.5ex} q) (1+q)
\times 8.36/{\rm sin}\hspace*{.5ex} i.}
\end{eqnarray}

Now, the procedure of constructing the model of the binary system is
the following one.

(1) First, we select an evolutionary model of HDE 226868 from Tab.
1.

(2) Second, we select the value of the coefficient $f_{\rm RL}$.

(3) Once the value of $f_{\rm RL}$ is assumed, we can solve the the
eqs. (3)$-$(4) for $i$ and $q$ and then get also the mass of the
compact component $M_{\rm X}$.

We do not have large freedom in selecting the value of $f_{\rm RL}$.
Earlier observations and analysis (Gies \& Bolton 1986a,b; C-N+09;
O+11) indicate that it has to be larger than 0.9 and, most likely,
not smaller than 0.95. Similarly, not all resulting values of the
inclination $i$ are acceptable. The same observations and analysis
indicate that it cannot be substantially different from about
30$^{\rm o}$ (O+11 got $i \sim 27^{\rm o}$).

Models of the binary system, obtained by taking evolutionary
supergiant models from Tab.1 and selecting the value of $f_{\rm RL}$
in the range 0.95 - 1 are listed in Tab. 2.

The last two lines of this table (corresponding to the solution
proposed by O+11) cannot describe realistic models, since they
require too small value of $f_{\rm RL}$ (let us note that parameters
suggested by O+11 imply $f_{\rm RL}$ $\approx$ 0.935, but they use
much smaller value for the mass of the optical component). Taken at
face value, Tab. 2 imply the mass of the black hole in the range 10
to 23 M$_{\odot}$. However, if one remembers what was said about the
realistic supergiant models (section 2.2) and if one requires
additionally that $f_{\rm RL}$ is $\ge$ 0.95, then one gets the
range 13 - 23 M$_{\odot}$. I consider this as the probable range for
the mass of the black hole in the system.

The most likely model is perhaps the one typed in boldface (second
line in Tab. 2). It corresponds to the Roche lobe filling factor
$f_{\rm RL}$ $\sim$ 0.965 and the inclination $i \sim 30^{\rm o}$. I
may add that, if one looks for the model with mass ratio $q$ equal
2.78 (as implied by an emission lines component discussed by Gies et
al. 2003), then (assuming still $M_{\rm opt}$ = 27.2 M$_\odot$) the
required inclination is $i \sim 46^{\rm o}$ (rather too high) and
the required value of the Roche lobe filling factor $f_{\rm RL}$
$\sim$ 0.92 (rather too small).

\begin{table}

\centering
 \begin{minipage}{75mm}
  \caption{Parameters of the selected evolutionary models of the
   binary system HDE 226868/Cyg X-1. The most likely model is the one typed in boldface (it corresponds to the mass of the optical component
$M_{\rm opt}$ equal 27 M$_\odot$, the Roche lobe filling factor
$f_{\rm RL}$ equal 0.965 and the inclination $i$ equal $30^{\rm
o}$).}

\vbox{
\begin{tabular}{|r|r|r|r|l|l|l|}
\hline
&&&&&&\\
\multicolumn{1}{|c|}{$d$}&\multicolumn{1}{|c|}{$T_{\rm
e}$}&\multicolumn{1}{|c|}{log$L$}&\multicolumn{1}{|c|}{$M_{\rm
opt}$}&
\multicolumn{1}{|c|}{$f_{\rm RL}$}&\multicolumn{1}{|c|}{$i$}&\multicolumn{1}{|c|}{$M_{\rm x}$}\\
\multicolumn{1}{|c|}{[kpc]}&\multicolumn{1}{|c|}{[10$^3$
K]}&\multicolumn{1}{|c|}{[L$_{\odot}$]}&\multicolumn{1}{|c|}{[M$_{\odot}$]}&&\multicolumn{1}{|c|}{[$^{\rm
o}$]}&\multicolumn{1}{|c|}{[M$_{\odot}$]}\\
&&&&&&\\
\hline

{\bf 1.86}&{\bf 28.0}&{\bf 5.309}&{\bf 27.2}&0.95&34.4&13.1\\
&&&&{\bf 0.965}&{\bf 29.3}&{\bf 15.8}\\
&&&&0.98&25.0&19.3\\
1.98&30.5&5.513&35.4&0.92&33.8&15.6\\
&&&&0.95&24.5&22.9\\
1.75&25.5&5.094&20.7&0.98&38.3&10.0\\
&&&&1.00&31.0&12.7\\
1.75&28.0&5.257&25.5&0.91&37.0&11.7\\
&&&&0.93&29.5&15.1\\
&&&&0.95&23.7&20.0\\
1.86&31.0&5.352&29.4&0.80&31.9&14.9\\
&&&&0.83&21.8&24.1\\
\hline

\end{tabular}}

\vspace{4mm}
{\footnotesize NOTES:\vspace{2mm}\\
$f_{\rm RL}$ denotes the Roche lobe fill-out factor; $i$ denotes the
inclination of the binary orbit; other symbols have their usual meanings\\
} \vspace{0mm}
\end{minipage}
\end{table}

\section{The effects of possible rapid rotation of the mass gainer at the end of the mass transfer}

\begin{table}

\centering
 \begin{minipage}{75mm}
  \caption{Present masses of the optical component for the models with lowered hydrogen content.The corresponding evolutionary tracks are shown in Fig. 3}

\vbox{
\begin{tabular}{|c|c|c|c|c|}
\hline
&&&&\\
\multicolumn{1}{|c|}{$X_{\rm 0}$}&\multicolumn{1}{|c|}{$M_{\rm
opt}$/M$_{\odot}$}&\multicolumn{1}{|c|}{acceptable?}&\multicolumn{1}{|c|}{$M_{\rm
opt}$/M$_{\odot}$}&\multicolumn{1}{|c|}{acceptable?}\\
&\multicolumn{1}{|c|}{(CN+09)}&&\multicolumn{1}{|c|}{(O+11)}&\\
&&&&\\
\hline

0.70&27.2&Y&29.4&Y\\

0.65&25.6&Y&28.0&Y\\

0.60&24.1&Y&26.4&Y\\

0.55&21.3&N&24.8&Y\\

\hline

\end{tabular}}

\vspace{4mm}
{\footnotesize NOTES:\vspace{2mm}\\
$X_{\rm 0}$ denotes the hydrogen content established as a result of
the rotation induced mixing at the end of the mass transfer; $M_{\rm
opt}$ denotes the present mass of the optical component for the
models corresponding to C-N+09 and O+11 solutions, respectively;
Y denotes acceptable and N non-acceptable models from the evolutionary point of view - see the text\\
} \vspace{0mm}
\end{minipage}
\end{table}

According to the analysis by C-N+09, the fit of the optical spectrum
of HDE 226868 is better for the model atmospheres with the hydrogen
content $X$ equal about 0.55 than for the solar abundances. This is
not a precise determination (as indicated by the authors) but it
suggests that HDE 226868 may have, at present, a decreased hydrogen
content. The reasons for such hydrogen depletion are not difficult
to find. In the course of its binary evolution, the system probably
underwent a major mass transfer, during which HDE 226868 was a mass
gainer. Accretion could spin up the star to very fast rotation. The
rotation could, in turn, induce large scale mixing which might even
homogenize the star. In this way, both the internal layers which
were partially hydrogen depleted due to earlier evolution of the
mass gainer and the newly dumped layers which were partially
hydrogen depleted due to earlier evolution of the mass donor were
mixed and homogenized. As a result, the hydrogen content of the
outer layers of HDE 226868 might be now lower that normal for the
Population I star ($X$ = 0.7). For the more extensive discussion of
the role of mass-transfer, rotation and mixing in the binary stars
evolution the reader is referred to de Mink et al. (2013).

The present rotation of HDE 226868 ($v$ sin $i$ $\approx$ 100 km/s,
C-N+09) is not very fast for an O type star. Such rotation does not
influence significantly the stellar structure. However, fast
rotation in the past could leave a permanent imprint in the form of
the lowered hydrogen content of the star. This would have
significant evolutionary consequences. It is difficult to estimate
quantitatively the amount of the accreted matter, its chemical
composition and the efficiency of the mixing. However, the overall
effect may be, very roughly, judged by parameterizing it with the
initial (i.e. after the mass transfer) hydrogen content in HDE
226868. This is a very crude approach, but it may give an idea about
the role of the processes mentioned above.

For this reason, I calculated the evolutionary tracks for stars with
the initial hydrogen content equal $X_{\rm 0}$ = 0.65, 0.60 and
0.55. The way of fitting the tracks was the same as for models
described in section 2.2 (I required that the luminosity, effective
temperature and the flux of the stellar wind will match the observed
values). The tracks were fitted to both solutions: C-N+09 and O+11.
The resulting tracks are shown in Fig. 3 (the tracks for $X_{\rm 0}$
= 0.70 are added for the comparison). The resulting values of the
present mass of HDE 226868 are listed in Tab. 3. As might be
expected, this mass decreases with the decreasing hydrogen content.
This reflects the well known fact that hydrogen depleted models are
more luminous and it is possible to match the observed luminosity
for the smaller mass. The smallest value of the present mass (21.3
M$_\odot$) is obtained for $X_{\rm 0}$ = 0.55 fitted to C-N+09
solution. This value would agree (within the existing uncertainties)
with the estimate of O+11. However, this model is not realistic, as
it represents the star which is in very fast post-main sequence
phase of evolution (shell hydrogen burning phase $-$ see Fig. 3d).
This model is increasing its radius at a rate of 1 \% in 50 years.

\begin{figure*}
\begingroup
   \def \A#1{\epsfxsize=0.48\hsize \epsfbox{#1}}
   \hbox to \hsize{\A{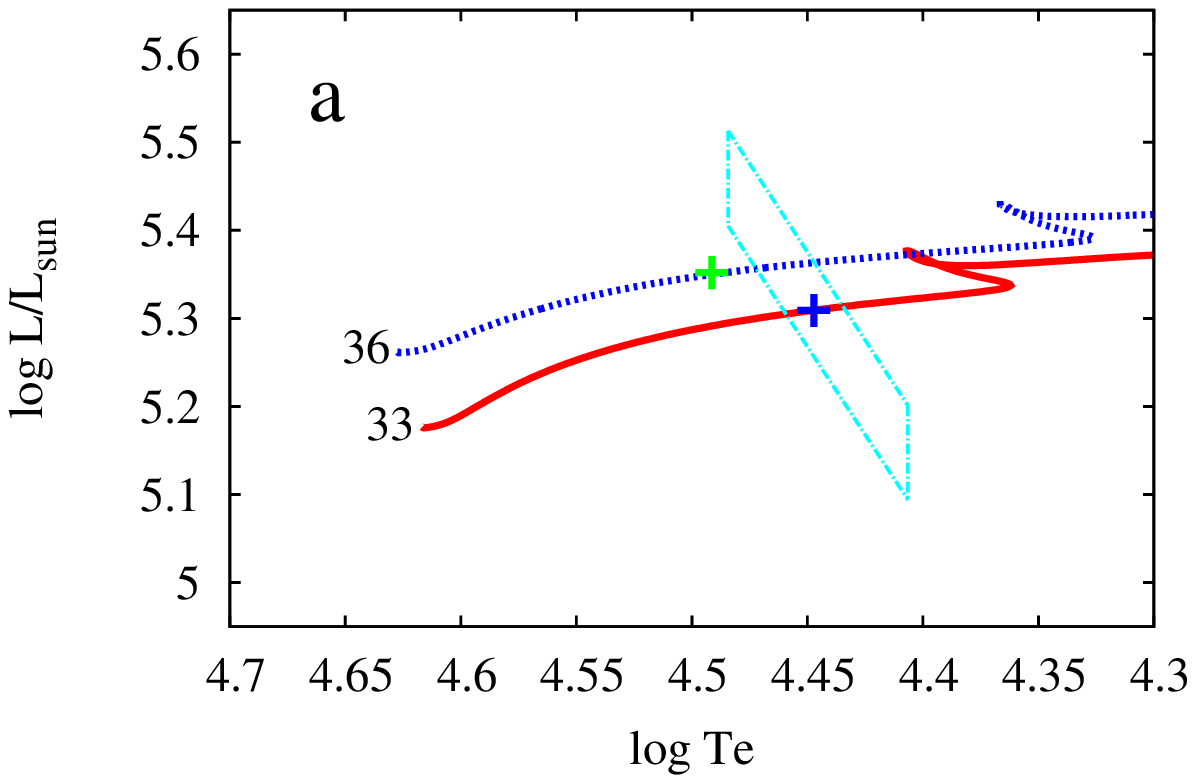}\hfil\A{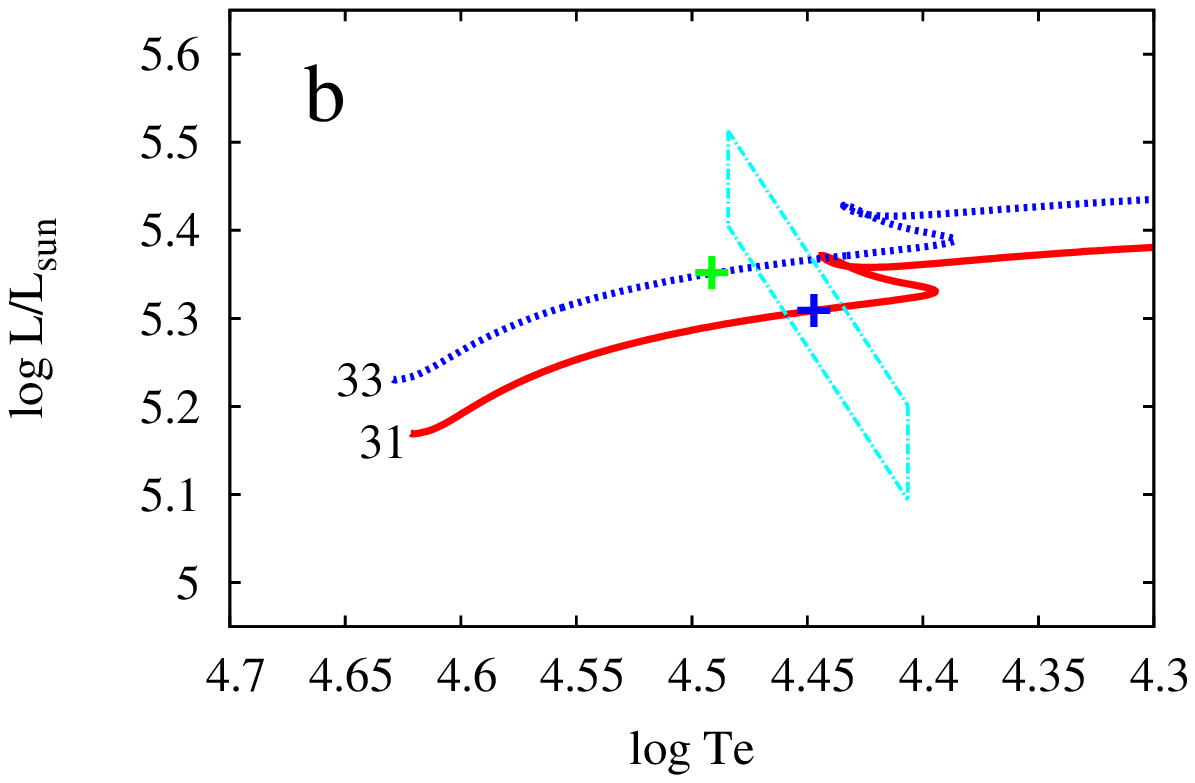}}
   \hbox to \hsize{\A{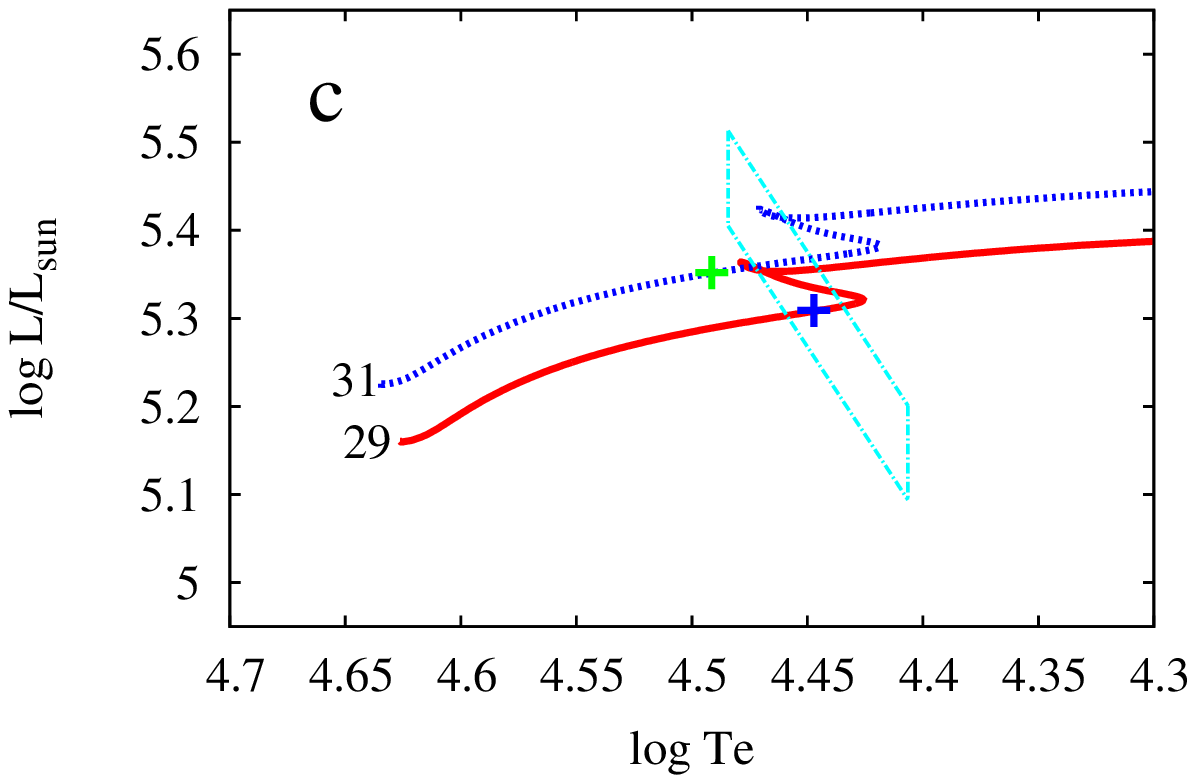}\hfil\A{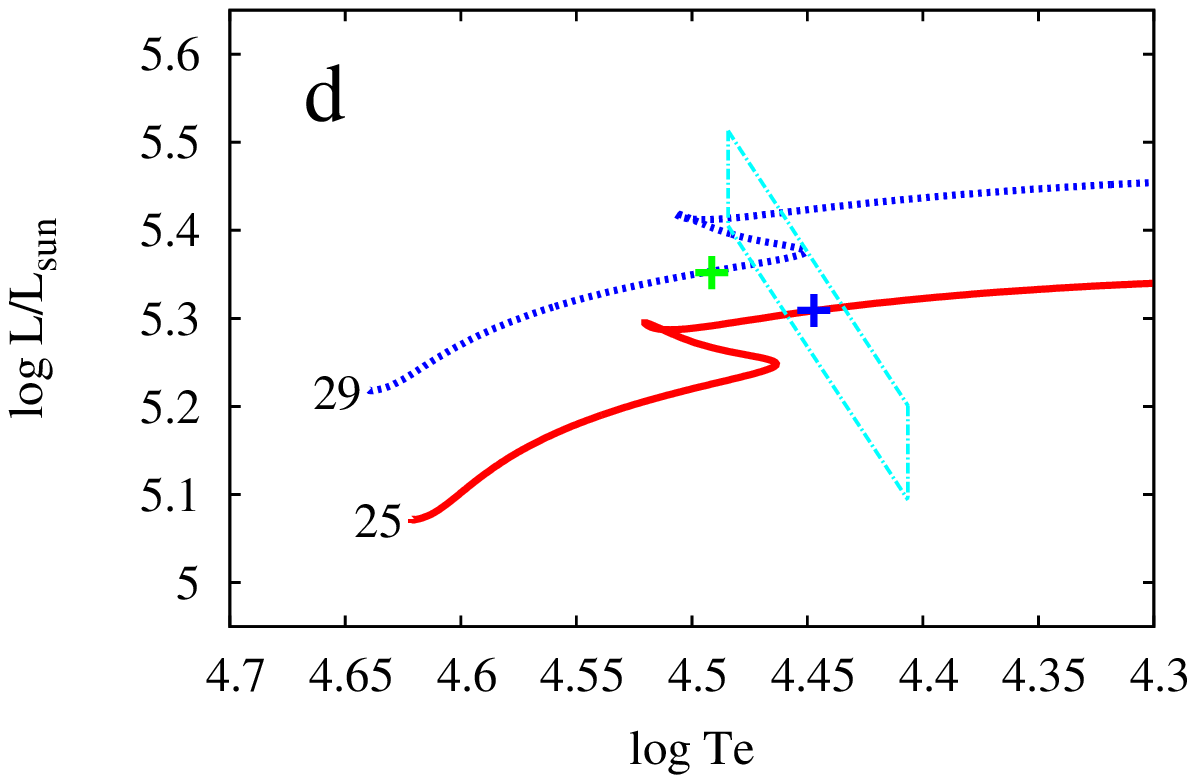}}
\endgroup
      \caption[h]{The evolutionary tracks in the H-R diagram for the
      different initial hydrogen content: (a) $X_{\rm 0} = 0.70$, (b) $X_{\rm 0} =
0.65$, (c) $X_{\rm 0} = 0.60$ and (d) $X_{\rm 0} = 0.55$. The tracks
are labeled with the initial mass of the star (in solar units).
Other symbols are the same as in Fig. 2. The second wiggle in each
evolutionary track (change of the direction of the evolution from
left to right) marks the end of the core hydrogen burning and the
beginning of the fast evolutionary phase.}
     \label{f3}
    \end{figure*}

Such rapid increase would have consequences for the flux of the
stellar wind (increase), the X-ray luminosity (increase) and the
orbital period of the system (change depends on the model of the
wind). None of these changes is seen during the forty years of
observations of HDE 226868, indicating that the star is in slow,
nuclear phase of its evolution. Therefore, we have to label this
solution with a letter N (model non-acceptable from the evolutionary
point of view) in Tab. 3. If we consider only models that are
acceptable (those labeled with Y in Tab. 3), then the smallest value
of the present mass of HDE 226868 is equal to 24.1 M$_\odot$. This
is not much smaller than the value 27.2 M$_\odot$ selected in the
earlier discussion.

\section{Conclusions}

(1) The present mass of HDE 226868 is, most likely, in the range 25
to 35 M$_\odot$ (estimate based on the evolutionary models). The
most likely value is 27 M$_\odot$ (corresponding to the central
values of R+11 distance and C-N+09 effective temperature).

(2) The mass of Cyg X$-$1 (the black hole) is, probably, in the
range 13 to 23 M$_\odot$. The most likely value is perhaps 16
M$_\odot$ (corresponding to the mass of the optical component
$M_{\rm opt}$ equal 27 M$_\odot$, a Roche lobe filling factor
$f_{\rm RL}$ equal 0.965 and an inclination $i$ equal $30^{\rm o}$).

(3) The value of the present mass of HDE 226868 as low as suggested
by O+11 (19 $\pm$ 2 M$_\odot$) is incorrect since it violates the
mass-luminosity relation for the the massive core hydrogen burning
stars (this relation was not incorporated in the iterative process
of O+11).

(4) There are observational indications that the hydrogen content of
HDE 226868 ($X \sim$ 0.55) might be lower than normal for a
population I star. This might be a result of the mass transfer, the
resulting fast rotation and the rotation induced mixing of the mass
gainer.

If this is confirmed (i.e. HDE 226868 has lowered hydrogen content),
then, according to my models, its present mass might be somewhat
lower ($\sim$ 24 M$_\odot$). However, it cannot be much lower (again
according to my models) if we require that the model corresponds to
the long-lasting evolutionary phase. If the lower hydrogen content
of HDE 226868 is confirmed, then a more detailed binary model
incorporating the effects of mass transfer and rotational mixing
will be desirable.

\section*{Acknowledgements}

I would like to thank A. Zdziarski for a careful reading of the
manuscript and for helpful comments. I would like also to thank the
anonymous referee for calling my attention to the necessity of
considering rotation as an important factor in the evolution of the
optical component. This work was partially supported by the Polish
Ministry of Science and Higher Education grant N203 581240 and by
the Polish National Science Center project 2012/04/M/ST9/00780.

\label{lastpage}
\end{document}